\documentclass[sigconf, screen]{acmart}

\AtBeginDocument{%
  }

\usepackage{adjustbox}    
\usepackage{multirow}
\setcopyright{none} 
\copyrightyear{2024}
\renewcommand\footnotetextcopyrightpermission[1]{}
\acmConference[]{}{}{}
\acmDOI{}
\acmISBN{}
\acmYear{}
\acmMonth{}
\setcopyright{none}
\begin{document}
%%
%% The "title" command has an optional parameter,
%% allowing the author to define a "short title" to be used in page headers.
\title{AEG: A Baremetal Framework for AI Acceleration via Direct Hardware Access in Heterogeneous Accelerators}
% \subtitle{\normalsize{ISCA 2025 Submission
%     \textbf{\#NaN} -- Confidential Draft -- Do NOT Distribute!!}}
% %%
%% The "author" command and its associated commands are used to define
%% the authors and their affiliations.
%% Of note is the shared affiliation of the first two authors, and the
%% "authornote" and "authornotemark" commands
%% used to denote shared contribution to the research.

\author{Hua Jiang}
\email{huaj@amd.com}
\orcid{0009-0008-3468-0098}
\affiliation{%
  \institution{AMD}
  \city{San Jose}
  \state{California}
  \country{USA}
}

\author{Sayan Mandal}
\email{saymanda@amd.com}
\orcid{0000-0003-2700-4629}
\affiliation{%
  \institution{AMD}
  \city{San Jose}
  \state{California}
  \country{USA}
}

\author{Brandon Kirincich}
\email{Brandon.Kirincich@amd.com}
\affiliation{%
  \institution{Rochester Institute of Technology}
  \city{Rochester}
  \state{New York}
  \country{USA}}

\author{Govind Varadarajan}
\email{govindarajulu.varadarajan@amd.com}
\affiliation{%
  \institution{AMD}
  \city{San Jose}
  \state{California}
  \country{USA}
}

% \author{}
% \affiliation{%
%   \institution{AMD}
%   \city{San Jose}
%   \state{California}
%   \country{USA}
% }

% \authorsaddresses{Corresponding author: Hua Jiang,
% \href{mailto:huaj@amd.com}{huaj@amd.com};}

%%
%% By default, the full list of authors will be used in the page
%% headers. Often, this list is too long, and will overlap
%% other information printed in the page headers. This command allows
%% the author to define a more concise list
%% of authors' names for this purpose.

%%
%% The abstract is a short summary of the work to be presented in the
%% article.
\begin{abstract}
This paper introduces a unified, hardware-independent baremetal runtime architecture designed to enable high-performance machine learning (ML) inference on heterogeneous accelerators, such as AI Engine (AIE) arrays, without the overhead of an underlying real-time or general-purpose operating system. Existing edge-deployment frameworks, such as TinyML, often rely on real-time operating systems (RTOS), which introduce unnecessary complexity and performance bottlenecks. To address this, our solution fundamentally decouples the runtime from hardware specifics by flattening complex control logic into linear, executable Runtime Control Blocks (RCBs). This "Control as Data" paradigm allows high-level models, including Adaptive Data Flow (ADF) graphs, to be executed by a generic engine through a minimal Runtime Hardware Abstraction Layer (RHAL). We further integrate Runtime Platform Management (RTPM) to handle system-level orchestration (including a lightweight network stack) and a Runtime In-Memory File System (RIMFS) to manage data in OS-free environments. We demonstrate the framework's efficacy with a ResNet-18 image classification implementation. Experimental results show 9.2$\times$ higher compute efficiency (throughput per AIE tile) compared to Linux-based Vitis AI deployment, 3--7$\times$ reduction in data movement overhead, and near-zero latency variance (CV~$=0.03\%$). The system achieves 68.78\% Top-1 accuracy on ImageNet using only 28 AIE tiles compared to Vitis AI's 304 tiles, validating both the efficiency and correctness of this unified bare-metal architecture.
\end{abstract}

%%
%% The code below is generated by the tool at http://dl.acm.org/ccs.cfm.
%% Please copy and paste the code instead of the example below.
%%
%\begin{CCSXML}
%<ccs2012>
% <concept>
%  <concept_id>00000000.0000000.0000000</concept_id>
%  <concept_desc>Do Not Use This Code, Generate the Correct Terms for Your Paper</concept_desc>
%  <concept_significance>500</concept_significance>
% </concept>
% <concept>
%  %<concept_id>00000000.00000000.00000000</concept_id>
%  <concept_desc>Do Not Use This Code, Generate the Correct Terms for Your Paper</concept_desc>
%  <concept_significance>300</concept_significance>
% </concept>
% <concept>
%  %<concept_id>00000000.00000000.00000000</concept_id>
%  <concept_desc>Do Not Use This Code, Generate the Correct Terms for Your Paper</concept_desc>
%  <concept_significance>100</concept_significance>
% </concept>
% <concept>
 % <concept_id>00000000.00000000.00000000</concept_id>
%  <concept_desc>Do Not Use This Code, Generate the Correct Terms for Your Paper</concept_desc>
%  <concept_significance>100</concept_significance>
% </concept>
%</ccs2012>
%\end{CCSXML}

%\ccsdesc[500]{Do Not Use This Code~Generate the Correct Terms for Your Paper}
%\ccsdesc[300]{Do Not Use This Code~Generate the Correct Terms for Your Paper}
%\ccsdesc{Do Not Use This Code~Generate the Correct Terms for Your Paper}
%\ccsdesc[100]{Do Not Use This Code~Generate the Correct Terms for Your Paper}

%%
%% Keywords. The author(s) should pick words that accurately describe
%% the work being presented. Separate the keywords with commas.
\keywords{Adaptive Data Flow (ADF),AI Acceleration,Bare metal Computing,Edge AI,Heterogeneous Accelerators,runtime control blocks}
%% A "teaser" image appears between the author and affiliation
%% information and the body of the document, and typically spans the
%% page.
% \begin{teaserfigure}
%   \includegraphics[width=\textwidth]{sampleteaser}
%   \caption{Seattle Mariners at Spring Training, 2010.}
%   \Description{Enjoying the baseball game from the third-base
%   seats. Ichiro Suzuki preparing to bat.}
%   \label{fig:teaser}
% \end{teaserfigure}

%\received{20 February 2007}
%\received[revised]{12 March 2009}
%\received[accepted]{5 June 2009}

%%
%% This command processes the author and affiliation and title
%% information and builds the first part of the formatted document.
\renewcommand\footnotetextcopyrightpermission[1]{}
\maketitle
\pagestyle{plain}

\section{Introduction}
The growing deployment of edge systems has intensified demand for low-latency, energy-efficient AI inference on heterogeneous accelerators. Platforms such as the AMD Versal Adaptive Compute Acceleration Platform (ACAP) integrate AIE vector processor arrays that provide high throughput for signal-processing and machine-learning workloads \cite{zhuang2023high}. However, the end-to-end utilization of such accelerators is frequently bottlenecked by software stacks that assume an OS-mediated execution environment \cite{chen2019deep,shao2022edge}. While OS-based abstractions improve portability via compiler stacks such as TVM and embedded ML runtimes, they introduce non-trivial overheads from kernel crossings, scheduler latency, and memory subsystem effects \cite{chen2018tvm, david2021tensorflow, mcvoy1996lmbench}. In our experiments, Linux kernel transitions can lead to a 7$\times$ increase in latency for small (1 KB) transfers, a critical penalty for inference pipelines that require frequent tensor movement between memory and compute units \cite{mcvoy1996lmbench, li2007quantifying}.

Current support for end-to-end baremetal (OS-less) software stacks that can orchestrate heterogeneous AI accelerators without relying on an operating system remains limited. Prior work on baremetal ML deployment motivates this direction, yet often results in rigid, hardware-specific solutions that lack scalability \cite{kumar2025bare}.

This paper addresses this gap by presenting a unified, hardware-independent baremetal runtime architecture. Unlike traditional approaches that hard-code control logic for specific devices, our framework fundamentally decouples the execution model from hardware specifics through a "Control-as-Data" paradigm. We introduce runtime control blocks (RCBs), which flatten complex graph execution semantics into linear, executable data sequences. This allows a generic runtime engine to orchestrate operations across diverse hardware targets from AMD AIEs to emerging spatial architectures without recompilation or OS intervention.

The main contributions of this work are:
\begin{itemize}
\item \textbf{Control-as-Data for Baremetal Resource Decoupling:} We address software resource limitations inherent to baremetal environments by adopting a control-as-data mechanism. By encoding execution semantics into structured, hardware-agnostic command streams (runtime control blocks) that resemble the task instruction approach in VTA \cite{moreau2018vta}, we decouple the runtime from OS-level software dependencies and eliminate the need for device-specific control logic within the runtime core.
\item \textbf{Layered Abstraction Architecture:} We introduce a Runtime Hardware Abstraction Layer (RHAL) to isolate hardware heterogeneity via a minimal primitive interface, and a Runtime In-Memory File System (RIMFS) to provide unified, file-like data management in OS-free environments.
\item \textbf{System-Level Orchestration:} We present Runtime Platform Management (RTPM), a module that assumes the role of a system executive, handling global cache coherency, interrupt dispatching, and secure network connectivity for remote provisioning.
\item \textbf{Performance \& Portability:} We demonstrate that this architecture not only facilitates the seamless integration of high-level frameworks (such as ADF and PyTorch) but also eliminates user-kernel switching overheads, delivering superior latency determinism and throughput compared to OS-based baselines.
\end{itemize}

\section{Background}

\subsection{AI Engine Architecture}

The AIE in AMD Versal devices represents a spatial accelerator paradigm: a two-dimensional array of tiles, each integrating a very long instruction word (VLIW) processor with single instruction multiple data (SIMD) capabilities, local SRAM, and a configurable AXI4-Stream switch \cite{AMD_UG1079_2025_2, AMD_WP506_2022, AMD_WP539_2025}. Unlike GPU architectures that rely on hardware-managed caches, AIE requires software to explicitly choreograph data movement via distributed direct memory access (DMA) engines using global memory I/O (GMIO), making it inherently suited to dataflow execution but demanding precise orchestration of transfers and buffer management \cite{AMD_UG1079_2025_2}.

\subsection{Existing AIE Deployment Approaches}

\textbf{Production runtime.} Vitis AI is the current state-of-the-art framework for deploying neural networks on AMD AIE arrays \cite{AMD_VitisAI_2024}. It provides quantization, compilation, and runtime (VART) through a Linux-based driver stack. While Vitis AI offers mature tile mapping and scheduling, it operates through kernel-mediated control: applications request driver cooperation, allocate memory via kernel interfaces, and issue DMA commands via ioctl calls.

\textbf{Compiler infrastructures.} MLIR-AIE \cite{mliraie} and MLIR-AIR \cite{mlirair} target the AIE compute fabric through compiler-managed scheduling, while XTA \cite{xta_special_session} pipelines graphs across system-on-chip (SoC) compute units. These solutions optimize compute placement but remain constrained by OS-level scheduling overhead.

\textbf{Edge ML frameworks.} General-purpose stacks such as TVM \cite{chen2018tvm} and TinyML runtimes (TensorFlow Lite Micro, CMSIS-NN) \cite{david2021tensorflow,lai2018cmsis} bridge high-level models to target code but typically assume OS-managed execution contexts.

\subsection{Baremetal Computing}

OS-mediated runtimes introduce control-path overheads from system calls, context switches, and scheduler latency \cite{mcvoy1996lmbench,buttazzo1997hard}. For streaming inference with frequent small transfers, these fixed costs can dominate end-to-end latency. Baremetal execution eliminates user-kernel crossings, improving both latency and predictability (reduced jitter).

However, baremetal deployment on spatial accelerators like AIE presents unique challenges:
\begin{itemize}
\item \textbf{Missing platform services:} No file system for weights, no TCP/IP stack, no dynamic memory manager, all must be reimplemented in lightweight form.
\item \textbf{Dependency isolation:} Standard runtimes depend on libraries requiring system calls; a baremetal solution must be self-contained.
\item \textbf{Explicit orchestration:} The spatially distributed compute array requires manual coordination of DMA transfers, buffer lifetimes, and tile synchronization \cite{AMD_WP552_2023}.
\end{itemize}

This motivates our central question: \emph{Can we preserve high-level programmability while eliminating OS overhead?} The following section presents a unified baremetal architecture that adopts a "Control as Data" philosophy, encoding execution semantics into hardware-agnostic runtime control blocks (RCBs), while internalizing platform services through modular components (RHAL, RIMFS, RTPM).

\section{Methodology}

\subsection{System Overview}

We propose a unified baremetal software stack that drastically reduces the integration complexity required to deploy existing ML models in OS-free environments. The overarching design goal is to establish a streamlined deployment path that allows standard ML models to run directly on baremetal, effectively replacing heavy OS-provided services such as memory allocation, device configuration, and peripheral I/O with a compact, application-level platform layer.

To demonstrate the efficacy of this approach, we utilize the AIE array and the ADF framework as a representative case study. In this reference implementation, our framework retains full compatibility with the standard toolchain, specifically the compiled ADF graph artifacts that map kernels and streams onto the physical array \cite{AMD_UG1079_2025_2}. By directly ingesting these artifacts, the system provides the essential control and data-movement services needed to run complex, pre-existing ML models on the hardware with minimal adaptation overhead.

\subsection{System Architecture}

\begin{figure*}[t]
\centering
\includegraphics[width=\textwidth]{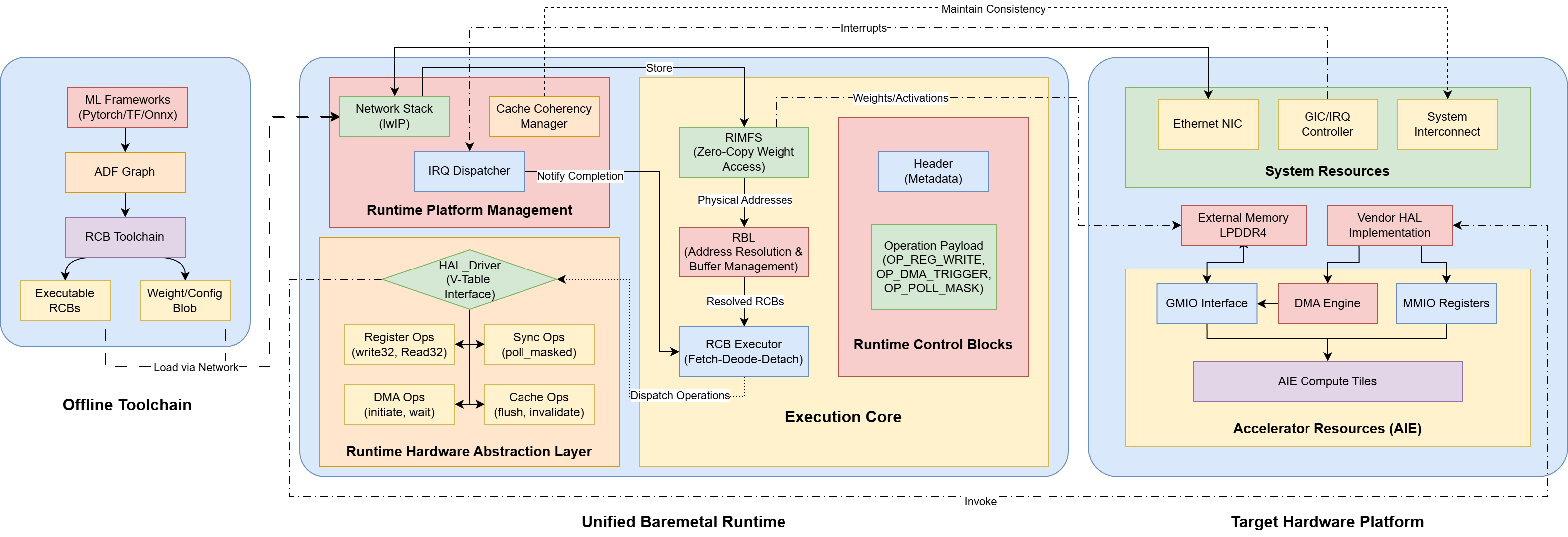}
\caption{Unified baremetal runtime architecture. The framework comprises three layers: (1) the Offline Toolchain (RCTC) that translates ADF graphs into executable RCBs and weight blobs; (2) the Unified Bare-Metal Runtime containing RTPM for platform services, RIMFS for zero-copy data management, RBL for address binding, and the RCB Executor; and (3) the Target Hardware Platform accessed through the RHAL abstraction layer.}
\Description{System architecture diagram showing the three-layer baremetal runtime: Offline Toolchain with RCTC, Unified Bare-Metal Runtime with RTPM/RIMFS/RBL/Executor/RHAL, and Target Hardware Platform with AIE tiles and memory.}
\label{fig:arch}
\end{figure*}

The proposed architecture adopts a "Control as Data" philosophy, in which complex software control logic is reduced to executable data structures rather than compiled into host machine code. This approach enables the runtime engine to remain generic and hardware-agnostic, while hardware-specific details are isolated in thin abstraction layers. Figure~\ref{fig:arch} conceptually decomposes the framework into a three-layer architecture: (i) the Offline Toolchain, (ii) the Unified Baremetal Runtime, and (iii) the Target Hardware Layer. The runtime layer comprises five core components: RCBs, RHAL, RIMFS, Runtime Binding Layer (RBL), and RTPM.

\subsubsection{RCB Toolchain and ADF Integration}

The RCTC bridges high-level ML frameworks and the baremetal runtime by performing forward translation of ADF computation graphs into symbolic RCBs. In AMD's programming model, an ADF graph is a network of kernels connected by data streams \cite{AMD_UG1079_2025_2}. RCTC converts these graph representations into executable RCBs while preserving the dataflow semantics.

The toolchain performs three key functions:
\begin{itemize}
\item \textbf{Forward translation:} Converting ADF computation graphs and layer representations into symbolic RCBs with hardware-agnostic operation sequences.
\item \textbf{Data packaging:} Flattening weights and configuration data into binary blobs suitable for RIMFS storage.
\item \textbf{Mapping generation:} Producing descriptors that map logical tensor IDs to physical requirements, resolved at runtime by the binding layer.
\end{itemize}

This design enables new models to be deployed by providing compiled ADF graph artifacts that are automatically translated into RCBs, without requiring modifications to the runtime or kernel drivers.

\subsubsection{Runtime Control Blocks}

The core of our execution model is the Runtime Control Block. An RCB is not executable code but a data structure containing a sequence of commands that encode the complete execution semantics of ML workloads. By representing control flow as data, we eliminate the need for the runtime to "know" the model structure, enabling a hardware-agnostic execution model.

Each RCB comprises:
\begin{itemize}
\item \textbf{Header:} Metadata including block type, size, and dependency information.
\item \textbf{Operation payload:} A structured sequence of low-level operations (e.g., \texttt{OP\_REG\_WRITE}, \texttt{OP\_DMA\_TRIGGER}, \texttt{OP\_POLL\_MASK}) that describe register writes, status reads, and programmed data-movement operations.
\end{itemize}

The RCB format is hardware-independent; target addresses within an RCB are either symbolic (resolved by the runtime binding layer) or relative, ensuring the same RCB structure can drive different accelerators when the RHAL layer is adapted. This design aligns with the AIE memory/transfer model, where GMIO connects the AIE array to global memory \cite{AMD_UG1079_2025_2}. By constructing command sequences and executing them directly in user space (baremetal), the framework avoids OS-mediated system calls and user-kernel crossings, improving control-path predictability.

\subsubsection{Runtime Hardware Abstraction Layer}

To achieve true hardware independence, we define a strict boundary between the generic runtime and vendor-specific hardware drivers via RHAL. This interface is implemented as a C-struct of function pointers (\texttt{hal\_driver\_t}), acting as a virtual function table (vtable) that encapsulates operations any hardware vendor must implement for integration.

RHAL categorizes hardware interactions into four fundamental primitives:
\begin{itemize}
\item \textbf{Register operations:} \texttt{write32}, \texttt{read32}, \texttt{write\_block} for configuring accelerator control registers (CSRs).
\item \textbf{DMA operations:} \texttt{initiate\_dma}, \texttt{wait\_dma} abstracting tensor data movement between system memory (RIMFS) and accelerator local memory via GMIO \cite{AMD_UG1079_2025_2}.
\item \textbf{Synchronization:} \texttt{poll\_register\_masked} handling handshakes between the host CPU and the accelerator.
\item \textbf{Cache coherency:} \texttt{flush\_cache}, \texttt{invalidate\_cache} for maintaining consistency between CPU caches and DRAM in ARM-based baremetal systems.
\end{itemize}

This design ensures that integrating a new accelerator requires only implementing this thin driver layer, without modifying core runtime logic. The runtime dynamically invokes hardware-specific routines at execution time without requiring compile-time knowledge of the target accelerator.

\subsubsection{Runtime In-Memory File System}

In bare-metal environments, standard file systems are unavailable or impose excessive overhead. RIMFS provides a read-only, flat-memory file abstraction for managing model weights, parameters, and metadata without reliance on persistent storage or OS-level file systems.

Key design features include:
\begin{itemize}
\item \textbf{Address mapping:} Maps file IDs (e.g., weight tensor IDs) to physical memory offsets, enabling direct data access.
\item \textbf{Zero-copy access:} Returns physical addresses directly to the DMA engine, allowing weight reads without CPU intervention or memory copying.
\item \textbf{Aligned allocation:} Provides regions suitable for GMIO transfers on Versal platforms \cite{AMD_UG1079_2025_2}.
\end{itemize}

On Versal platforms, AIE access to external memory is performed through GMIO. RIMFS tracks buffer ownership across receive/compute/send stages while exposing stable physical addresses to both the networking layer and AIE transfer configuration.

\subsubsection{Runtime Binding Layer}

The RBL serves as the execution coordinator between symbolic RCBs and physical hardware resources. Its responsibilities include:

\begin{itemize}
\item \textbf{Data binding:} Maps RCB symbolic inputs, outputs, and weights to physical memory locations in RIMFS, ensuring each RCB executes with correct data without embedding hardware-specific addresses.
\item \textbf{Address resolution:} Resolves symbolic buffer IDs and logical offsets at runtime, allocating buffers and computing physical/DMA addresses for RCB execution.
\item \textbf{Dependency and buffer management:} Tracks intermediate buffer usage across multiple RCBs in a pipeline, manages input/output handoffs between sequential or parallel RCBs, and maintains buffer lifetimes for efficient memory utilization.
\end{itemize}

RBL produces fully resolved, executable RCBs for the runtime executor while remaining decoupled from hardware operations. It works closely with the RCB executor and RHAL to enable hardware-agnostic execution.

\subsubsection{Runtime Platform Management}

While RHAL manages accelerator interactions, RTPM manages the broader system environment as a lightweight system executive. In the absence of an operating system, RTPM orchestrates global system resources through three critical functionalities:

\begin{itemize}
\item \textbf{Cache coherency management:} Manages interconnect consistency protocols to ensure data integrity between the host CPU, DMA engines, and accelerators.
\item \textbf{Asynchronous event handling:} Provides a unified interrupt service routine (ISR) dispatcher to handle hardware signals, error exceptions, and completion notifications, replacing standard OS interrupt stacks.
\item \textbf{Host connectivity and telemetry:} Integrates a lightweight TCP/IP stack (lwIP) for remote model loading, runtime control, and real-time performance telemetry \cite{dunkels2003full,dunkels2001design}. Each message includes a CRC-32 checksum (IEEE polynomial $0x04\mathrm{C}11\mathrm{DB}7$) for error detection \cite{Infineon_FCE_XMC_2016}.
\end{itemize}

RTPM enables the baremetal runtime to operate as a network-attached inference service, receiving inputs and returning results without OS-mediated I/O.

\subsubsection{Execution Flow}

\begin{figure}[t]
\centering
\includegraphics[width=\columnwidth]{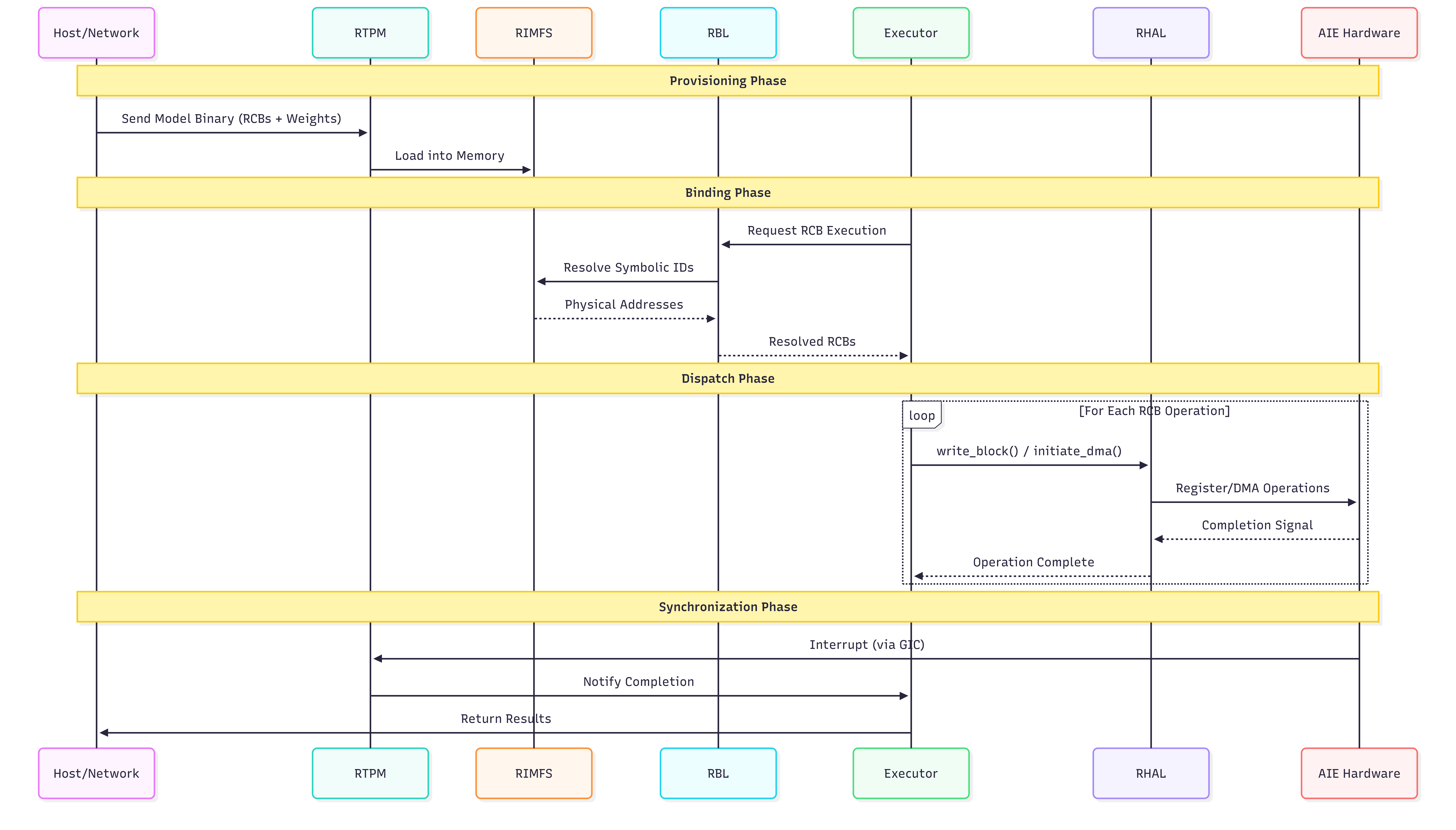}
\caption{Execution flow sequence showing the four phases: Provisioning (model loading via RTPM), Binding (symbolic-to-physical address resolution via RBL/RIMFS), Dispatch (RCB operation execution via RHAL), and Synchronization (interrupt-driven completion notification).}
\Description{Sequence diagram showing execution flow between Host, RTPM, RIMFS, RBL, Executor, RHAL, and AIE Hardware components across provisioning, binding, dispatch, and synchronization phases.}
\label{fig:exec-flow}
\end{figure}

The runtime execution follows a cyclic "Fetch-Decode-Dispatch" pattern coordinated across the architectural components, as illustrated in Figure~\ref{fig:exec-flow}:

\begin{enumerate}
\item \textbf{Provisioning:} RTPM receives the model binary (RCBs and weights) via Ethernet and loads them into RIMFS.
\item \textbf{Binding:} RBL parses the RCBs and resolves symbolic IDs (e.g., "Weight\_Tensor\_01") into physical addresses provided by RIMFS.
\item \textbf{Dispatch:} The executor iterates through RCB instructions, invoking RHAL primitives: \texttt{write\_block()} for computation configuration, \texttt{initiate\_dma()} for data movement with resolved physical addresses.
\item \textbf{Synchronization:} The runtime invokes \texttt{poll()} or waits for signals from RTPM's interrupt dispatcher to confirm task completion before proceeding to subsequent layers.
\end{enumerate}

This flow ensures deterministic execution while maintaining full compatibility with the ADF graph representation used in the AIE toolchain \cite{AMD_UG1079_2025_2}.

\subsection{ResNet-18 Case Study}

We evaluate the framework by deploying ResNet-18 for 1000-class ImageNet classification \cite{he2016deep, russakovsky2015imagenet}. The network is expressed as a pipelined ADF graph, with weights stored in external memory and activations streamed through the AIE kernels. The framework instantiates the graph, initializes parameter buffers (12.63~MB in our implementation), and orchestrates end-to-end inference from packet arrival to result transmission. This case study exercises (i) graph execution control, (ii) repeated external-memory transfers, and (iii) sustained networking I/O.

\subsection{Experimental Setup}

\textbf{Hardware platform.} Experiments were conducted on an AMD Versal AI Edge VEK280 evaluation kit (VE2802 device). The kit integrates ML-optimized AI Engines and external LPDDR4 memory (12~GB on the board) \cite{AMD_VEK280_kit_web}. We use GMIO as the primary mechanism to transfer tensors between external memory and the AIE array \cite{AMD_UG1079_2025_2}.

\textbf{Measurement procedure.} To isolate control-path and data-path effects, we measure three intervals: (1) input transfer time (external memory $\rightarrow$ AIE via GMIO), (2) kernel execution time (AIE compute), and (3) output transfer time (AIE $\rightarrow$ external memory via GMIO). Each experiment is repeated for $>1000$ iterations; we report summary statistics (mean and percentiles) and discard initial warm-up iterations to reduce cold-cache effects.

We include two microbenchmarks: (i) a pass-through kernel that performs no arithmetic (transfer-dominated), and (ii) a $64\times 64$ matrix multiplication kernel (combined transfer and compute). These kernels allow separating the impact of transfer orchestration from arithmetic intensity.

\textbf{Accuracy validation.} We validate the ResNet-18 deployment using 5000 randomly sampled images from the ILSVRC2012 validation set (50000 labeled images total) \cite{russakovsky2015imagenet}. Images are resized/cropped to $224\times 224$ and normalized using standard ImageNet preprocessing. We quantize inputs to INT8 and compute Top-1 and Top-5 accuracy.

\textbf{Linux-based AIE baseline.} We use Vitis AI as the state-of-the-art baseline for AIE deployment, as it represents the current production-grade runtime, and no other publicly available AIE inference frameworks exist. To compare against the standard OS-based AIE deployment, we evaluate ResNet-18 using Vitis AI 5.1 on the same VEK280 hardware running Linux \cite{AMD_VitisAI_2024}. We use a pre-trained PyTorch ResNet-18 model quantized through the Vitis AI quantization flow and deploy it using the Vitis AI Runtime (VART) API \cite{AMD_VitisAI_Tutorial_ResNet18}. Inference is executed on the neural processing unit (NPU) with preprocessing and postprocessing performed on the Arm CPU; only NPU inference time is included in the reported latency. Profiling indicates that Vitis AI utilizes 304 AIE tiles, while our baremetal deployment uses a 4$\times$7 grid (28 tiles). This difference enables compute efficiency comparisons that normalize for resource utilization.

\textbf{Kernel-user mode elimination.} A key architectural difference between the two AIE deployments is the control path. Vitis AI operates through the Linux kernel driver stack: user-mode applications request driver cooperation, perform memory allocation/attachment via kernel interfaces, communicate addresses through IOCtl calls, and rely on the kernel to issue DMA commands. Our baremetal approach eliminates these transitions: the application runs directly in privileged mode, allocates memory through direct C library APIs, and issues DMA commands without kernel mediation. This elimination of kernel crossings is the primary mechanism by which the baremetal framework achieves higher compute efficiency per tile.

\textbf{Data-path correctness.} To verify functional correctness of baremetal transfers and buffer addressing, we implement two additional tests: (i) a $64\times 64$ matrix multiplication kernel, and (ii) a small neural pipeline (Conv2D $\rightarrow$ ReLU $\rightarrow$ Softmax). Reference outputs are generated with NumPy using deterministic seeds and compared to AIE outputs at runtime.

\section{Results}

\subsection{Performance Analysis: Effect of OS-less Hardware Control}

A central objective of the proposed framework is to reduce control-path latency by executing device configuration and data-movement orchestration without OS-mediated system calls. We therefore benchmarked the latency of representative hardware operations implemented via our command-based control interface and compared them with functionally equivalent operations in a Linux-hosted deployment.

\textbf{Hardware control and transfer latency.} For transfers associated with a $64\times 64$ matrix workload, the proposed framework achieves a 3.3$\times$ reduction in per-operation overhead compared to Linux. For small-block transfers (1~KB), baremetal achieves a 7.0$\times$ reduction in per-transfer overhead. Table~\ref{tab:data-movement} reports relative overhead as a function of block size while holding the total transferred volume constant (100~MB). The relative benefit is most significant for small blocks, consistent with a control-dominated regime in which fixed per-transfer costs dominate end-to-end time.

\begin{table}[t]
\centering
\caption{Per-Transfer Overhead Reduction vs. Block Size (100~MB Total).}
\label{tab:data-movement}
\begin{tabular}{lr}
\toprule
\textbf{Block Size} & \textbf{Speedup (Baremetal vs. Linux)} \\
\midrule
1 KB & 7.0$\times$ \\
4 KB & 5.4$\times$ \\
16 KB & 3.0$\times$ \\
32 KB & 2.2$\times$ \\
\bottomrule
\end{tabular}
\end{table}

\textbf{End-to-end pipeline time.} We additionally measure the complete inference flow that includes (i) ADF graph instantiation, (ii) memory region initialization, and (iii) graph execution. The full pipeline achieves a 40$\times$ reduction in baremetal compared to Linux. This gap indicates that OS-mediated control and data movement can dominate end-to-end execution time when the application repeatedly configures transfers and synchronizes on fine-grained events.

\begin{figure}[t]
\centering
\includegraphics[width=0.9\columnwidth]{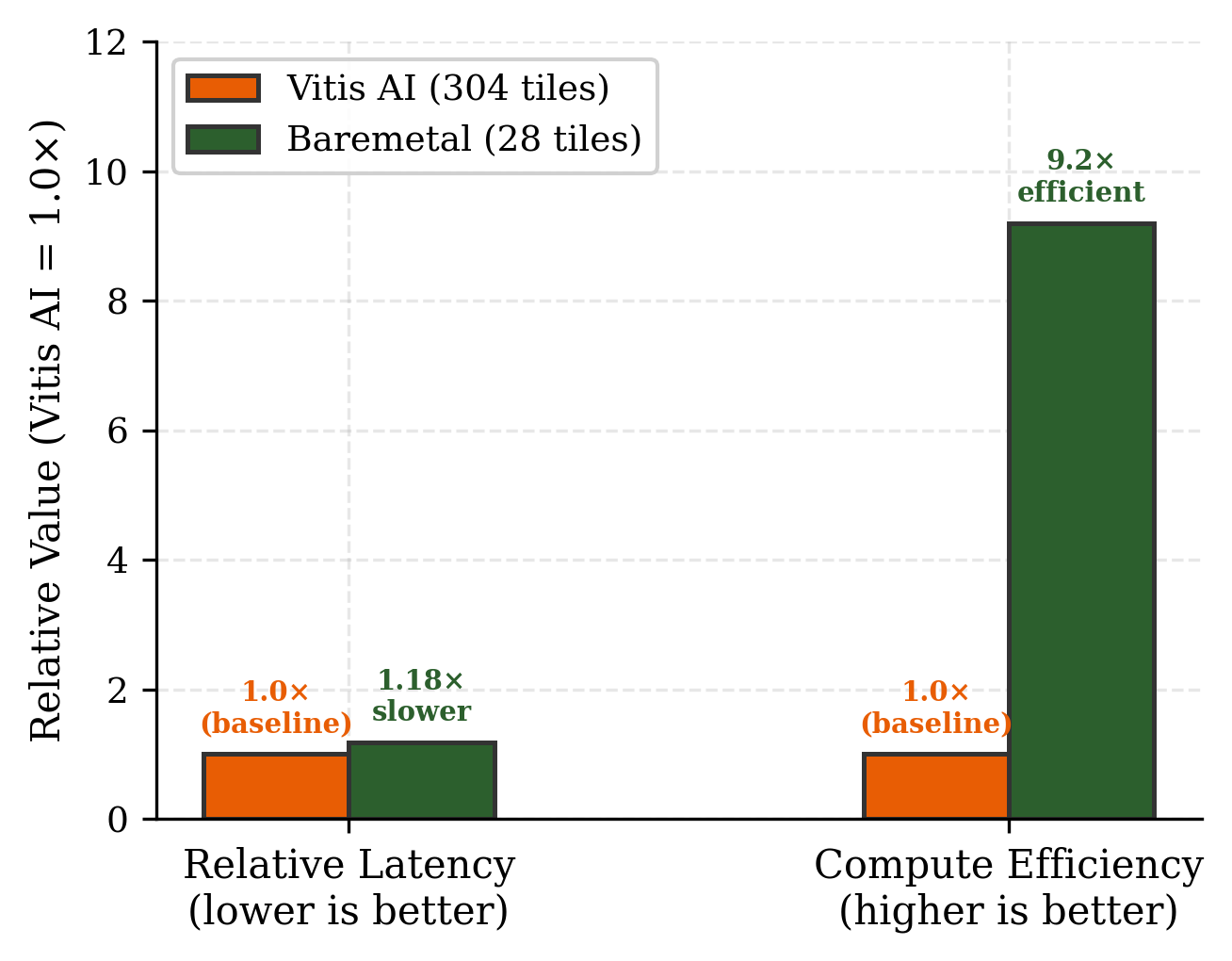}
\caption{ResNet-18 inference comparison: Baremetal vs.\ Vitis AI (304-tile Linux baseline). Left: relative latency (lower is better); baremetal achieves 1.18$\times$ baseline latency using only 28 tiles. Right: compute efficiency (higher is better); baremetal achieves 9.2$\times$ higher throughput per tile. This demonstrates that eliminating kernel-user mode transitions enables competitive latency with substantially fewer resources.}
\label{fig:inference-latency}
\end{figure}

\textbf{Inference latency and variability.} We quantify variability using the coefficient of variation (CV~$= \sigma/\mu$), a dimensionless measure of dispersion. The baremetal system exhibits CV~$=0.03\%$, compared to CV~$=0.63\%$ for the Linux-based Vitis AI deployment. Figure~\ref{fig:inference-latency} visualizes both the relative latency and compute efficiency between the two AIE deployments. The low variance observed in baremetal is consistent with eliminating OS scheduling effects from the critical path.

\subsection{Resource Utilization}

Removing the operating system reduces both non-model memory requirements and startup overhead.

\begin{itemize}
\item \textbf{Memory footprint.} Including model parameters, the complete baremetal image achieves $>$1.1$\times$ smaller non-volatile storage footprint and $>$2.7$\times$ smaller runtime memory compared to Linux (Yocto) baselines \cite{YoctoWiki_ImageSize_2011}. The majority of runtime memory is consumed by model weights, with minimal overhead for input/output buffers.
\item \textbf{Startup latency.} The system reaches a network-ready state approximately 350--745$\times$ faster than Linux (Yocto), which reports representative boot times on the order of tens of seconds with significant BIOS overhead \cite{YoctoWiki_BootTime_2011}. This reduces time-to-service by roughly two to three orders of magnitude.
\end{itemize}

\begin{table}[t]
\centering
\caption{Resource utilization improvement over Linux (Yocto-referenced baselines).}
\label{tab:resources}
\begin{tabular}{lr}
\toprule
\textbf{Metric} & \textbf{Baremetal vs.\ Linux} \\
\midrule
Image size (incl.\ weights) & $>$1.1$\times$ smaller \\
Runtime memory (incl.\ weights) & $>$2.7$\times$ smaller \\
Time to network-ready & $\sim$350--745$\times$ faster\footnotemark \\
\bottomrule
\end{tabular}
\end{table}

\footnotetext{Yocto boot-time example reports tens of seconds, with a significant portion attributable to BIOS initialization.}

\subsection{Accuracy, Efficiency, and Correctness}

\textbf{Accuracy and efficiency.} On 5000 images sampled from the ImageNet validation set, the baremetal AIE deployment achieves 68.78\% Top-1 accuracy and 88.22\% Top-5 accuracy using only 28 AIE tiles (4$\times$7 grid). The Linux-based Vitis AI deployment achieves 69.00\% Top-1 and 88.54\% Top-5 accuracy but utilizes 304 AIE tiles. Table~\ref{tab:accuracy} summarizes accuracy and compute efficiency. We define efficiency as throughput per tile: $(1/\text{latency}) / \text{tiles}$.

While Vitis AI achieves 1.18$\times$ lower raw latency than our baremetal implementation (due to its larger tile allocation and optimized tile mapping), the baremetal deployment achieves 9.2$\times$ higher per-tile efficiency. This indicates that Vitis AI's latency advantage comes primarily from utilizing $\sim$11$\times$ more compute resources rather than from more efficient execution. The baremetal approach demonstrates that eliminating kernel-user mode transitions enables competitive performance with substantially fewer resources, a critical consideration for power-constrained edge deployments.

\begin{table}[t]
\centering
\caption{ResNet-18 Inference: Baremetal vs.\ Vitis AI (ImageNet, 5{,}000 images). Efficiency is computed as throughput per AIE tile; higher values indicate better resource utilization.}
\label{tab:accuracy}
\begin{tabular}{lrr}
\toprule
\textbf{Metric} & \textbf{Baremetal (ours)} & \textbf{Vitis AI (Linux)} \\
\midrule
AIE Tiles & 28 & 304 \\
Top-1 Accuracy & 68.78\% & 69.00\% \\
Top-5 Accuracy & 88.22\% & 88.54\% \\
Relative Latency & 1.18$\times$ & 1.0$\times$ (baseline) \\
Compute Efficiency & 9.2$\times$ & 1.0$\times$ (baseline) \\
CV (Variability) & 0.03\% & 0.63\% \\
\bottomrule
\end{tabular}
\end{table}

\textbf{Data-path correctness.} We evaluated the functional correctness of the baremetal transfer path using two hand-written kernel tests. For XGEMM ($64\times 64$ matrix multiplication), all 4096 output elements matched the reference (100\%). For the neural pipeline (Conv2D $\rightarrow$ ReLU $\rightarrow$ Softmax), all nine outputs matched the reference (9/9). Across these tests, we observed no data corruption, indicating that the explicit buffer management and direct transfer orchestration are functionally reliable under the evaluated conditions.

% \begin{figure}[t]
% \centering
% \includegraphics[width=\columnwidth]{data_movement_overhead_relative.png}
% \caption{Relative per-transfer overhead under Linux (OS-mediated) compared to bare metal baseline (1.0$\times$) as a function of block size. The overhead reduction is largest for small transfers (7$\times$ at 1~KB), consistent with fixed per-transfer control overheads dominating in the OS-mediated configuration.}
% \label{fig:data-movement}
% \end{figure}

\section{Discussion}

\subsection{Baremetal Platform Support for AI Accelerators}

The results indicate that the proposed modular architecture comprising RCBs, RHAL, RIMFS, RBL, and RTPM can provide the core services required to deploy and execute AIE workloads at the edge without OS dependencies. In contrast to deployments that rely on a general-purpose OS or an RTOS to supply scheduling, memory allocation, and device I/O, the proposed framework integrates these services within a unified runtime stack. The measured runtime footprint ($>$2.7$\times$ smaller than Linux) and time-to-network-ready ($\sim$350--745$\times$ faster than Linux boot) suggest that practical platform support for AIE can be achieved within tight resource budgets. Beyond reducing memory and boot overhead, the single-binary deployment model reduces operational complexity: the system is immediately capable of inference following reset, without multi-stage OS initialization and service startup.

The "Control as Data" philosophy embodied by RCBs provides additional benefits: execution semantics are encoded in data structures rather than compiled code, enabling runtime introspection, debugging, and potential optimization without recompilation. The RHAL abstraction layer further ensures that the core runtime logic remains unchanged when targeting different accelerator variants.

\textbf{Reporting methodology.} Throughout this paper, we report performance metrics primarily as relative improvements over baseline systems rather than absolute values. This choice is deliberate: absolute latencies and memory footprints are highly dependent on specific hardware configurations, toolchain versions, clock frequencies, and silicon revisions that may differ across device generations and product variants. Reporting absolute values would anchor readers to a particular configuration that may not reflect production deployments or future hardware iterations. In contrast, the relative improvements we report, which isolate the effect of eliminating OS-mediated control paths, represent an architectural insight that generalizes across configurations. The 3--7$\times$ reduction in per-transfer overhead and 9.2$\times$ efficiency improvement reflect fundamental differences in control-path design rather than artifacts of a specific measurement setup.

\subsection{Implications of RCB and ADF Integration}

The RCB-based execution model, combined with direct integration of compiled computational graphs, is the key mechanism by which the framework preserves programmability in a baremetal setting. Our successful ResNet-18 deployment, with accuracy matching the Linux-based Vitis AI baseline, demonstrates that graph IRs can serve as a stable intermediate representation for portability across execution environments. For example, when using ADF, the RCTC toolchain translates the graph IR into RCBs, while RBL performs runtime binding of symbolic references to physical addresses in RIMFS. However, the framework itself is not coupled to ADF; ADF is simply one representative IR among others that can be supported.

From a developer perspective, this design shifts effort away from per-model low-level integration (custom drivers and ad hoc buffer handling) and toward maintaining a reusable execution substrate. The practical implication is that model variation, including changes in topology and layer ordering, does not require re-implementing platform services or modifying RHAL implementations, as long as the resulting graph conforms to the supported programming interface.

\subsection{Latency Predictability and Control-Path Costs}

The performance results highlight that the primary advantage of the proposed approach is not only competitive mean latency but also substantially reduced variability. The near-zero inference variance (CV~$=0.03\%$, compared to CV~$=0.63\%$ for Vitis AI) is consistent with removing scheduler-related timing variation from the execution path. This property is consequential for closed-loop and real-time settings where worst-case latency and jitter directly affect stability and quality of service.

The RCB-based execution model contributes to this predictability: each RCB encodes a deterministic sequence of operations, and the RHAL primitives provide direct hardware access without OS-mediated indirection. The data-movement experiments further suggest that the dominant penalty in the OS-mediated configuration is a fixed per-transfer cost, rather than bandwidth limitations. The speedup is most significant for 1~KB transfers (7$\times$) and decreases as transfer size increases to 32~KB (2.2$\times$), indicating that amortization reduces the relative impact of control overheads. For neural inference, where intermediate activations are frequently moved in structured but relatively small blocks, the fixed-cost regime can be a limiting factor. These results motivate optimization strategies that either (i) reduce the number of transfers (fusion, buffering, and batching) or (ii) reduce the control cost per transfer (direct RHAL-based control as in this work).

\subsection{Compute Efficiency and Tile Utilization}

A key finding of this work is that eliminating kernel-user mode transitions enables dramatically higher \emph{compute efficiency}, defined as throughput per AIE tile, rather than merely improving raw latency. The baremetal deployment achieves 9.2$\times$ higher efficiency than Vitis AI while using only 28 tiles (4$\times$7 grid) compared to Vitis AI's 304 tiles. Despite using approximately 10$\times$ fewer compute resources, the baremetal system achieves inference latency within 18\% of that of the Linux-based deployment.

This efficiency advantage stems from the architectural differences in control flow. Vitis AI operates through a conventional Linux driver stack:
\begin{enumerate}
\item User-mode application requests kernel driver cooperation
\item Driver performs memory allocation and attachment via kernel interfaces
\item IOCtl calls communicate memory addresses to the driver
\item Kernel issues DMA commands on behalf of the application
\end{enumerate}
In contrast, our baremetal approach eliminates these transitions:
\begin{enumerate}
\item Application runs directly in privileged mode (no kernel/user separation)
\item Memory allocation uses direct C library APIs to reserve regions
\item No attachment or IOCtl overhead
\item DMA commands issued directly by the application
\end{enumerate}

The tile count difference arises because Vitis AI uses a closed-source, production-grade compiler with proprietary tile-placement and routing algorithms. Our implementation cannot replicate the exact tile mapping since these optimizations are not publicly documented. The tile utilization data (304 tiles) was obtained through execution profiling rather than architectural documentation. Theoretically, if our baremetal framework adopted an equivalent tile-mapping strategy, inference latency could match or exceed that of Vitis AI while preserving the determinism benefits demonstrated in this work.

It is important to note that our primary contribution is demonstrating that eliminating kernel-user mode switches enables efficient accelerator utilization, not optimizing tile-level scheduling. The 9.2$\times$ efficiency improvement validates this architectural hypothesis. Future work could integrate more sophisticated tile mapping while retaining the baremetal control path.

Despite the absolute latency difference, the baremetal approach achieves significantly lower variance (CV~$=0.03\%$ vs.\ CV~$=0.63\%$), confirming that OS-related timing jitter remains a factor even in well-optimized Linux deployments. For applications where worst-case latency guarantees are paramount, the baremetal approach offers advantages that complement raw throughput optimization.

\subsection{Networking Integrity, Threat Model, and Extension Points}

The networking layer is designed to minimize overhead and to detect transfer errors. The current design uses CRC-32 to detect accidental corruption and does not provide confidentiality or cryptographic authentication. This is an explicit trade-off: adding encryption and message authentication would increase compute and/or latency and would require key management. For deployments that require adversarial resistance (e.g., untrusted networks), a natural extension is to incorporate a lightweight authenticated encryption scheme (e.g., AES-GCM) or to terminate TLS at a trusted gateway and keep the device on a protected network segment. The appropriate option depends on the threat model and the acceptable overhead.

\subsection{Limitations and Future Work}

While the current evaluation demonstrates substantial gains, several limitations remain:

\begin{itemize}
\item \textbf{Generality of platform services.} The RTPM module implements the subset of platform functions needed for the evaluated workloads. Broader device support (additional peripherals, multiple network interfaces, or storage) may require extending RTPM while maintaining determinism.
\item \textbf{RHAL portability validation.} While the RHAL abstraction is designed for hardware independence, the current implementation has been validated only on the AIE platform. Porting to other accelerators would exercise the abstraction layer's generality.
\item \textbf{Preprocessing overhead.} End-to-end throughput is limited primarily by preprocessing. Moving preprocessing onto the accelerator, expressing it as RCBs, or overlapping preprocessing with inference via pipelining are likely to yield significant improvements.
\item \textbf{Robustness under load.} The reported correctness checks show no corruption in the evaluated runs; however, longer-duration stress tests with varying packet sizes, concurrent requests, and fault injection would better characterize RIMFS and RTPM reliability in production conditions.
\item \textbf{Multi-model and multi-tenant operation.} The current design targets a single deployed graph. Supporting concurrent graphs or dynamic admission control would require extending RBL with policies for memory partitioning, execution arbitration, and isolation.
\end{itemize}

\section{Conclusion}

This paper presented a unified, hardware-agnostic baremetal runtime architecture for AI accelerators, comprising RCBs that encode execution semantics as data, RHAL for portable accelerator interaction, RIMFS for zero-copy data management, and RTPM for system-level orchestration. By adopting a "Control as Data" philosophy and directly integrating with compiled ADF graphs, the framework eliminates OS dependencies while maintaining toolchain compatibility. Experimental evaluation demonstrates practical deployment footprints ($>$2.7$\times$ smaller runtime memory, $\sim$350--745$\times$ faster boot time vs.\ Linux) alongside substantial efficiency improvements: 9.2$\times$ higher compute efficiency (throughput per tile) compared to Linux-based Vitis AI, 3--7$\times$ reduction in per-transfer overhead, and near-zero latency variance (CV~$=0.03\%$). These results confirm that eliminating kernel-user mode transitions materially improves both efficiency and predictability in accelerator-based inference, enabling competitive performance with substantially fewer compute resources (28 vs.\ 304 AIE tiles). The modular architecture provides a practical path toward resource-efficient, latency-predictable inference on heterogeneous edge accelerators, with RHAL enabling future portability to additional hardware targets.

%%
%% The acknowledgments section is defined using the "acks" environment
%% (and NOT an unnumbered section). This ensures the proper
%% identification of the section in the article metadata, and the
%% consistent spelling of the heading.

%%
%% The next two lines define the bibliography style to be used, and
%% the bibliography file.
\bibliographystyle{ACM-Reference-Format}
\bibliography{sample-base}

@String{BIT = "{BIT}" }

@String{Computing = "Computing" }

@String{Computer = "{IEEE} Computer" }

@String{Springer = "Springer-Verlag" }

@inproceedings{zhuang2023high,
  title={High performance, low power matrix multiply design on acap: from architecture, design challenges and dse perspectives},
  author={Zhuang, Jinming and Yang, Zhuoping and Zhou, Peipei},
  booktitle={2023 60th ACM/IEEE Design Automation Conference (DAC)},
  pages={1--6},
  year={2023},
  organization={IEEE}
}

@article{chen2019deep,
  title={Deep learning with edge computing: A review},
  author={Chen, Jiasi and Ran, Xukan},
  journal={Proceedings of the IEEE},
  volume={107},
  number={8},
  pages={1655--1674},
  year={2019},
  publisher={IEEE}
}

@inproceedings{shao2022edge,
  title={Edge-rt: Os support for controlled latency in the multi-tenant, real-time edge},
  author={Shao, Wenyuan and Ye, Bite and Wang, Huachuan and Parmer, Gabriel and Ren, Yuxin},
  booktitle={2022 IEEE Real-Time Systems Symposium (RTSS)},
  pages={1--13},
  year={2022},
  organization={IEEE}
}

@inproceedings{chen2018tvm,
  title={$\{$TVM$\}$: An automated $\{$End-to-End$\}$ optimizing compiler for deep learning},
  author={Chen, Tianqi and Moreau, Thierry and Jiang, Ziheng and Zheng, Lianmin and Yan, Eddie and Shen, Haichen and Cowan, Meghan and Wang, Leyuan and Hu, Yuwei and Ceze, Luis and others},
  booktitle={13th USENIX Symposium on Operating Systems Design and Implementation (OSDI 18)},
  pages={578--594},
  year={2018}
}

@article{david2021tensorflow,
  title={Tensorflow lite micro: Embedded machine learning for tinyml systems},
  author={David, Robert and Duke, Jared and Jain, Advait and Janapa Reddi, Vijay and Jeffries, Nat and Li, Jian and Kreeger, Nick and Nappier, Ian and Natraj, Meghna and Wang, Tiezhen and others},
  journal={Proceedings of machine learning and systems},
  volume={3},
  pages={800--811},
  year={2021}
}

@inproceedings{mcvoy1996lmbench,
  title={Lmbench: Portable tools for performance analysis.},
  author={McVoy, Larry W and Staelin, Carl and others},
  booktitle={USENIX annual technical conference},
  pages={279--294},
  year={1996},
  organization={San Diego, CA, USA}
}

@inproceedings{li2007quantifying,
  title={Quantifying the cost of context switch},
  author={Li, Chuanpeng and Ding, Chen and Shen, Kai},
  booktitle={Proceedings of the 2007 workshop on Experimental computer science},
  pages={2--es},
  year={2007}
}

@article{kumar2025bare,
  title={Bare-Metal RISC-V+ NVDLA SoC for Efficient Deep Learning Inference},
  author={Kumar, Vineet and Li, Yike and Shanker, Shreejith and John, Deepu and others},
  journal={arXiv preprint arXiv:2508.16095},
  year={2025}
}

@techreport{AMD_WP552_2023,
  title        = {AI Engine Programming: A Kahn Process Network Evolution},
  institution  = {{Advanced Micro Devices, Inc.}},
  type         = {White Paper},
  number       = {WP552},
  year         = {2023},
  month        = jul,
  url          = {https://docs.amd.com/r/en-US/wp552-ai-kpn},
  note         = {Revision 1.0; Release date 2023-07-20},
}

@manual{AMD_UG1079_2025_2,
  title        = {AI Engine Kernel and Graph Programming Guide},
  organization = {{Advanced Micro Devices, Inc.}},
  number       = {UG1079},
  year         = {2025},
  month        = nov,
  note         = {Version 2025.2 English; Release date 2025-11-26},
  url          = {https://docs.amd.com/r/en-US/ug1079-ai-engine-kernel-coding},
}

@techreport{AMD_WP506_2022,
  title        = {AI Engines and Their Applications},
  institution  = {{Advanced Micro Devices, Inc.}},
  type         = {White Paper},
  number       = {WP506},
  year         = {2022},
  month        = dec,
  note         = {Revision 1.2 English; Release date 2022-12-16},
  url          = {https://docs.amd.com/v/u/en-US/wp506-ai-engine},
}

@techreport{AMD_WP539_2025,
  title        = {System-Level Benefits of the Versal Platform},
  institution  = {{Advanced Micro Devices, Inc.}},
  type         = {White Paper},
  number       = {WP539},
  year         = {2025},
  month        = feb,
  note         = {Revision 1.2.1 English; Release date 2025-02-13},
  url          = {https://docs.amd.com/v/u/en-US/wp539-versal-system-level-benefits},
}

@article{lai2018cmsis,
  title={Cmsis-nn: Efficient neural network kernels for arm cortex-m cpus},
  author={Lai, Liangzhen and Suda, Naveen and Chandra, Vikas},
  journal={arXiv preprint arXiv:1801.06601},
  year={2018}
}

@book{buttazzo1997hard,
  title={Hard real-time computing systems: predictable scheduling algorithms and applications},
  author={Buttazzo, Giorgio C},
  year={1997},
  publisher={Springer}
}

@misc{AMD_VEK280_kit_web,
  title        = {AMD Versal{TM} AI Edge Series VEK280 Evaluation Kit},
  organization = {{Advanced Micro Devices, Inc.}},
  url          = {https://www.amd.com/en/products/adaptive-socs-and-fpgas/evaluation-boards/vek280.html},
  urldate      = {2025-12-15}
}

@inproceedings{dunkels2003full,
  title={Full TCP/IP for 8-bit architectures},
  author={Dunkels, Adam},
  booktitle={Proceedings of the 1st international conference on Mobile systems, applications and services},
  pages={85--98},
  year={2003}
}

@article{dunkels2001design,
  title={Design and Implementation of the lwIP TCP/IP Stack},
  author={Dunkels, Adam},
  journal={Swedish Institute of Computer Science},
  volume={2},
  number={77},
  year={2001}
}

@misc{Infineon_FCE_XMC_2016,
  title        = {{FCE}: Flexible {CRC} Engine: {XMC}{TM} microcontrollers},
  organization = {{Infineon Technologies AG}},
  howpublished = {Training material (PDF slides)},
  year         = {2016},
  month        = sep,
  note         = {Dated September 2016; PDF filename: Infineon-IP\_FCE\_XMC4-TR-v01\_01-EN},
  url          = {https://www.infineon.com/dgdl/Infineon-IP_FCE_XMC4-TR-v01_01-EN.pdf?fileId=5546d4624ad04ef9014b0780ca5b2265},
  urldate      = {2025-12-15}
}

@inproceedings{he2016deep,
  title={Deep residual learning for image recognition},
  author={He, Kaiming and Zhang, Xiangyu and Ren, Shaoqing and Sun, Jian},
  booktitle={Proceedings of the IEEE conference on computer vision and pattern recognition},
  pages={770--778},
  year={2016}
}

@article{russakovsky2015imagenet,
  title={Imagenet large scale visual recognition challenge},
  author={Russakovsky, Olga and Deng, Jia and Su, Hao and Krause, Jonathan and Satheesh, Sanjeev and Ma, Sean and Huang, Zhiheng and Karpathy, Andrej and Khosla, Aditya and Bernstein, Michael and others},
  journal={International journal of computer vision},
  volume={115},
  number={3},
  pages={211--252},
  year={2015},
  publisher={Springer}
}

@online{YoctoWiki_BootTime_2011,
  title        = {Linux Kernel/Boot Time},
  organization = {{Yocto Project}},
  year         = {2011},
  month        = jun,
  note         = {Wiki page; last edited 2011-06-23. Permanent revision (oldid=2501).},
  url          = {https://wiki.yoctoproject.org/wiki/index.php?title=Linux_Kernel/Boot_Time&oldid=2501},
  urldate      = {2025-12-15}
}

@online{YoctoWiki_ImageSize_2011,
  title        = {Linux kernel/Image Size},
  organization = {{Yocto Project}},
  year         = {2011},
  month        = jun,
  note         = {Wiki page; last edited 2011-06-27 21:53. Permanent revision (oldid=2527).},
  url          = {https://wiki.yoctoproject.org/wiki/index.php?title=Linux_kernel/Image_Size&oldid=2527},
  urldate      = {2025-12-15}
}

@inproceedings{xta_special_session,
  title={Special session: XTA: Open source extensible, scalable and adaptable tensor architecture for AI acceleration},
author={Ravikumar V Chakaravarthy, Hua Jiang},
  booktitle={2020 IEEE 38th International Conference on Computer Design (ICCD)},
  year={2020}
}

@inproceedings{mlirair,
  title={From Loop Nests to Silicon: Mapping AI Workloads onto AMD NPUs with MLIR-AIR},
author={Erwei Wang and Samuel Bayliss and Andra Bisca and Zachary Blair and Sangeeta Chowdhary, Kristof Denolf and Jeff Fifield and Brandon Freiberger and Erika Hunhoff and Phil James-Roxby, Jack Lo and Joseph Melber and Stephen Neuendorffer, Eddie Richter and Andre Rosti and Javier Setoain and Gagandeep Singh and Endri Taka and Pranathi Vasireddy and Zhewen Yu and Niansong Zhang and Jinming Zhuang},
  booktitle={	arXiv:2510.1487},
  year={2025}
}

@inproceedings{mliraie,
  title={2025. MLIR-AIE.},
author={Advanced Micro Devices.},
  booktitle={ https://xilinx.github.io/mlir-aie/},
  year={2025}
}

@article{moreau2018vta,
  title={VTA: An open hardware-software stack for deep learning},
  author={Moreau, Thierry and Chen, Tianqi and Zipkin, Luis and Wan, Ziheng and Lian, Ruiqi and Zhang, Huaibei andhree, Joshua and Perkins, Runjie and Ceze, Luis},
  journal={arXiv preprint arXiv:1807.04188},
  year={2018}
}

@misc{AMD_VitisAI_2024,
  title={{Vitis AI 5.1 User Guide}},
  author={{AMD}},
  year={2024},
  howpublished={\url{https://vitisai.docs.amd.com/en/latest/docs/install/install.html}},
  note={Accessed: 2025}
}

@misc{AMD_VitisAI_Tutorial_ResNet18,
  title={{Vitis AI Tutorial: Custom ResNet-18 Deployment on NPU}},
  author={{AMD}},
  year={2024},
  howpublished={\url{https://github.com/Xilinx/Vitis-AI-Tutorials/tree/5.1/Tutorials/public_VitisAI-NPU-Custom-ResNet18-Deployment}},
  note={Accessed: 2025}
}

%%
%% If your work has an appendix, this is the place to put it.
\pagebreak
\appendix

\section{Detailed Benchmark Data}

\subsection{Matrix Multiplication Kernel (1000 Iterations)}

Table~\ref{tab:mm-breakdown} presents the relative speedup breakdown for the matrix multiplication kernel benchmark, comparing baremetal execution against Linux.

\begin{table}[h]
\centering
\caption{Matrix Multiplication Kernel: Baremetal Speedup vs.\ Linux (1000 iterations)}
\label{tab:mm-breakdown}
\begin{tabular}{lr}
\toprule
\textbf{Metric} & \textbf{Speedup} \\
\midrule
Input Data Transfer & 3.0$\times$ \\
Output Data Transfer & 3.7$\times$ \\
Total Data Movement & 3.3$\times$ \\
Kernel Execution & 1.0$\times$ \\
\bottomrule
\end{tabular}
\end{table}

\subsection{Passthrough Kernel (1000 Iterations)}

Table~\ref{tab:pt-breakdown} presents the relative speedup for the passthrough kernel, which measures pure data movement overhead.

\begin{table}[h]
\centering
\caption{Passthrough Kernel: Baremetal Speedup vs.\ Linux (1000 iterations)}
\label{tab:pt-breakdown}
\begin{tabular}{lr}
\toprule
\textbf{Metric} & \textbf{Speedup} \\
\midrule
Data Movement & 3.1$\times$ \\
Kernel Execution & 2.6$\times$ \\
Total Execution & 3.0$\times$ \\
\bottomrule
\end{tabular}
\end{table}

\section{Memory Layout}

The baremetal executable memory is allocated across three primary regions. The text section (executable code and constants) comprises the largest portion, followed by initialized global variables in the data section. Runtime buffers are allocated for model weights (which dominate memory usage), input feature maps, and output feature maps. The compact buffer allocation reflects the zero-copy design of RIMFS, which avoids intermediate buffering overhead.

\end{document}